\documentclass[superscriptaddress, amsmath, amssymb, aps, rmp, floatfix, twocolumn]{revtex4-2}
\pdfoutput=1
\usepackage[utf8]{inputenc}
\usepackage[T1]{fontenc}
\usepackage{babel}
\usepackage{graphicx}
\usepackage{amsfonts}
\usepackage{amssymb}
\usepackage{amsmath}
\usepackage{amsthm}
\usepackage{mathtools}
\usepackage{physics}
\usepackage[normalem]{ulem}
\usepackage{blindtext}
\usepackage{dsfont}

\def\id{{\rm 1\kern-.22em l}}

\usepackage{mathtools}
\usepackage{braket}

\usepackage{graphicx}
\usepackage[format=plain,justification=centerlast]{caption}
\usepackage[format=plain,justification=centerlast]{subcaption}
\usepackage{bm}
\usepackage[dvipsnames]{xcolor}

\usepackage{natbib}
\setcitestyle{square,numbers}
\usepackage[colorlinks]{hyperref}

\begin{document}
\title{Relational Quantum Mechanics is Still Incompatible 
with Quantum Mechanics}

\author{Jay Lawrence}
\affiliation{Department of Physics and Astronomy,
Dartmouth College, Hanover, NH 03755, USA}

\author{Marcin Markiewicz}
\affiliation{Institute of Theoretical and Applied Informatics, Polish Academy of Sciences, ul. Baltycka 5, 44-100 Gliwice, Poland}

\author{Marek \.Zukowski}
\affiliation{International Centre for Theory of Quantum Technologies (ICTQT),
University of Gdansk, 80-308 Gdansk, Poland}


\begin{abstract}
We showed in a recent article (Lawrence et. al., 2023, Quantum \textbf{7},
1015),
that relative facts (outcomes), a central concept in Relational Quantum 
Mechanics, are inconsistent with Quantum Mechanics. We proved this 
by constructing a Wigner-Friend type sequential measurement scenario on 
a Greenberger-Horne-Zeilinger (GHZ) state of three qubits, and making 
the following assumption: ``if an interpretation of quantum theory 
introduces some conceptualization of outcomes of a measurement, then 
probabilities of these outcomes must follow the quantum predictions 
as given by the Born rule.'' Our work has been criticized by 
Cavalcanti, Di Biagio, and Rovelli (CDR).  In this note we show that 
their critique is invalid, and that their specific arguments raise
questions of principle. 
\end{abstract}

\maketitle


Relational Quantum Mechanics (RQM) holds that an entangling interaction 
between an observer (Alice) and the object of study (the system $S$) 
constitutes a measurement, as a result of which the observer 
``realizes'' a relative fact - one of the possible  outcomes. This outcome
exists relative to Alice, but it 
does not exist for another observer (Bob) who does not interact with 
Alice or the system itself.  We demonstrated in \cite{Lawrence22}, 
hereafter called LMZ, that such relative facts as defined within RQM 
lead to a GHZ-like contradiction with quantum mechanics.  The proof is 
mathematically identical, but physically distinct, from that derived 
by \cite{Mermin.90} in his demonstration that non-contextual hidden 
variables contradict quantum mechanics. Our proof employes a sequential 
measurement scenario of the Wigner-and-Friend type, as is often seen in
RQM arguments.  Specifically, Alice ($A$) performs measurements on the 
system ($S$) consisting of three qubits prepared in a GHZ state.  By 
measuring each qubit separately, she obtains three relative facts 
$\mathcal{A}_1$, $\mathcal{A}_2$, and $\mathcal{A}_3$. Then Bob performs 
complementary measurements on the joint system $S \otimes A$ (also, 
effectively, a three-qubit system), obtaining relative facts  
$\mathcal{B}_1$, $\mathcal{B}_2$, and $\mathcal{B}_3$.  Bob's 
measurements result in a GHZ state of the compound system 
$S \otimes A \otimes B$.  Correlations inherent in this state may be
expressed as constraints on products of relative facts.
These are identical to the four constraints derived by Mermin.  We list 
them here for ease of reference in the coming discussion:
\begin{eqnarray}
    \mathcal B_1 \mathcal B_2 \mathcal B_3&=&1 , \label{GHZ-CONTR-1}\\
    \mathcal B_1 \mathcal A_2 \mathcal A_3&=&-1 , \label{GHZ-CONTR-2}\\
    \mathcal A_1 \mathcal B_2 \mathcal A_3&=&-1 , \label{GHZ-CONTR-3}\\
 \mathcal A_1 \mathcal A_2 \mathcal B_3&=&-1 . \label{GHZ-CONTR-4}
 \end{eqnarray}
Incidentally, to see the contradiction, simply equate the product of 
left-hand sides with the product of right-hand sides: $\mathcal A_1^2 \mathcal A_2^2 \mathcal A_3^2 \mathcal B_1^2 \mathcal B_2^2 
    \mathcal B_3^2 = -1.$
No set of assignments $A_1 = \pm 1$, etc., can satisfy this set of 
constraints.

Our paper has been criticized by \cite{CBR}, hereafter called CDR.  
Their main points of criticism are based on a reformulation of our 
proof, made ostensibly for the sake of clarity, but this is an 
illusion.  While the reformulation produces the same constraint 
equations as ours, the physical situation is different, both as 
viewed by Bob, or by a Wigner who measures Bob’s relative facts. 
The equivalence suggested by equations and diagrams does not apply 
to the physical situations described in the two scenarios.  As a 
result, the critique does not, in fact, apply to our proof.  Let us 
describe two specific points in the reformulation that show this.

\medskip
{\bf Point 1:} CDR replace each of Alice and Bob with 
three observers, called respectively $A_1$, $A_2$, $A_3$, and 
$B_1$, $B_2$, $B_3$, so that there is a separate
observer for each relative fact. They 
claim (erroneously) that we do the same, and later in the paper 
they criticize us for it, saying: ``{\it One cannot 
however simply define an observer $B = B_1 \otimes B_2 \otimes B_3$ 
relative to which constraint (1) holds} [{\it i.e.}, $B_1B_2B_3 = 1]$, 
{\it if there is no interaction involving those three systems after 
their measurements take place.''}  CDR resolve this (manufactured) 
problem by having a Wigner ``measure’’ the three relative facts.
 Why then cannot a single Bob do the same?  We see no
reason against this in the written postulates of RQM. 
\cite{Rovelli.21}.
\begin{itemize} 
\item{The objection  quoted above does not apply to 
our proof,  because a single observer 
($B$) makes all three measurements.}
\end{itemize}

\medskip
{\bf Point 2:} CDR reformulate $B$'s
overall measurement strategy, 
and this leads to their main critique, which they consider to be 
fatal. Instead of letting $B$ simply measure the three relative facts 
of the system $S \otimes A$, CDR imagine four experiments, one for 
each constraint equation. In the first experiment, each $B_m$
(m $= 1,2,3)$ reverses the measurement of $A_m$, and then performs 
his own (complementary) measurement on the qubit $S_m$.  Wigner has 
access to the three relative facts, and these must 
satisfy Eq. (1). In a second experiment, only one of the $B$'s, say 
$B_1$, reverses $A_1$’s measurement and performs his own 
measurement on the qubit $S_1$. The other $B$'s do nothing. This 
leaves three relative facts, one relative to each of $B_1$, $A_2$, 
and $A_3$, to which Wigner has access, and these must satisfy 
constraint (2).  Similar experiments result in constraints (3) and 
(4). The point made by CDR is that any one of the constraints may 
be satisfied, but the four constraints cannot be satisfied 
together, because there is no single observer for whom all 
six measurements are made in the same experiment.

\begin{itemize}
\item{But the above CDR scenario is not what we do, and so the 
critique does not apply to our scenario.  Ours involves a single
experiment in which $A$ and $B$ each perform just three 
commuting measurements, obtaining a total of six relative facts.   
Owing to GHZ correlations, and given our assumption 
regarding measurement outcomes as} stated in the abstract, all 
four constraints must be satisfied.
\end{itemize}

\noindent The above points illustrate that two different scenarios,
representing two different physical situations, can both lead to
the same four constraint equations.  One scenario (LMZ) proves
that relative facts are inconsistent with quantum mechanics; the
other (CDR) does not.

A particularly glaring distortion of our presentation 
 regarding point 2 
involves the $B$s' reversals of the $A$s' measurements. As we
mentioned above, each $B_m$ reverses the entangling measurement of 
$A_m$ with the qubit $S_m$, thus restoring the initial states of 
$S$ and $A$.  This raises the question of the meaning of the 
``realization'' of a relative fact by an observer.  For example, 
if $A$ is human, then she must see the process by which she 
realizes a relative fact as nonunitary.  When $B$ reverses the
unitary process of her entanglement with a qubit, what becomes 
of the nonunitary part as seen by $A$?  Is this undone or not?

Suppose, on the other hand, that the observer $A$ is simply 
another qubit?  What does it mean for the ``observer qubit'' 
to ``realize'' a relative fact?  We argue that such a 
``realization'' is a meaningless concept in quantum mechanics;
it is not possible to test this hypothesis by any experiment.

There are further inconsistencies and misleading 
remarks in the CDR paper which relate to such fundamental issues, 
but which are incidental to our point in this paper.  Two of us expand on these points in a separate paper \cite{Markiewicz23}. 
A principle concern is the following:  In the last page of their comment, 
CDR speak of ``invoking’’ the new {\it cross perspective link} postulate introduced by \cite{Adlam.22}:
{\em In a scenario where some observer Alice measures
a variable $V$ of a system $S$, then provided that Alice does not undergo any interactions which
destroy the information about $V$ stored in Alice’s physical variables, if Bob subsequently measures the physical variable representing Alice’s information about the variable $V$, then Bob’s
measurement result will match Alice’s measurement result.}
This postulate profoundly changes the 
nature of relative outcomes in RQM.  It makes $A$’s relative outcomes 
potentially accessible to any other observer making the prescribed 
measurement. However, this allows one to apply the 
reasoning of \cite{Zukowski.21} (steps 2 and 3) involving the notion 
of counterfactual values to show the internal inconsistency of such 
a reformulation of RQM.

In summary, CDR reformulate our 
proof to reproduce the statistics of measurement outcomes, but this 
does not reproduce the physics.  Their critique is based in large part 
upon their reformulation, and does not apply to what we actually did.  
Their critique needlessly employs multiple observers while a single 
observer would suffice.  It replaces our single experiment consisting of 
sequential measurements, by their four (mutually exclusive) experimental 
situations and employs a reversal operation which appears to us as 
ill-defined within the RQM  postulates.

MZ is  supported by  Foundation for Polish Science (FNP), IRAP project ICTQT, contract no. 2018/MAB/5, co-financed by EU  Smart Growth Operational Programme.
%
\end{document}